\documentclass[aps,prb,twocolumn, superscriptaddress,nobibnotes]{revtex4-1}

\usepackage{graphicx}
\usepackage{pstricks}
\usepackage{amsmath}
\usepackage{gensymb}
\usepackage{amsfonts}
\usepackage{dcolumn}
\usepackage{bm}
\usepackage{pdfpages}

\makeatletter
\AtBeginDocument{\let\LS@rot\@undefined}
\makeatother

\begin{document}
	
	\title{Effect of non-local correlations on the electronic structure of LiFeAs}
	\author{Karim Zantout}
	\email{zantout@itp.uni-frankfurt.de}
	\affiliation{Institut f\"ur Theoretische Physik, Goethe-Universit\"at 
		Frankfurt, 60438 Frankfurt am Main, Germany}		
	\author{Steffen Backes}
	\affiliation{CPHT, CNRS, Institut Polytechnique de Paris, F-91128 Palaiseau, France}
	\author{Roser Valent\'i}
	\affiliation{Institut f\"ur Theoretische Physik, Goethe-Universit\"at Frankfurt, 60438 Frankfurt am Main, Germany}
	
\begin{abstract}
We investigate the role of non-local correlations in LiFeAs by
exploring an ab-initio-derived  multi-orbital Hubbard model 
for LiFeAs via the Two-Particle Self-Consistent (TPSC) approach. The multi-orbital formulation of TPSC 
approximates the irreducible interaction vertex to be an orbital-dependent constant, which is self-consistently
determined from local spin and charge sum rules. Within this approach, we disentangle the contribution of local and non-local correlations in LiFeAs and show that in the local approximation one recovers the dynamical-mean field theory (DMFT) result.
The comparison of our theoretical results to most recent angular-resolved photoemission spectroscopy (ARPES) and de-Haas van Alphen (dHvA)
data shows that non-local correlations in LiFeAs are decisive to describe the measured 
spectral function $A(\vec k,\omega)$, Fermi surface and scattering rates.
These findings underline the importance of non-local correlations and benchmark
different theoretical approaches for iron-based superconductors.
\end{abstract}
	
	\maketitle
	
	{\it Introduction.-}
The nature of the electronic structure in 
iron-based superconductors has been intensively scrutinized since
their discovery in 2008\cite{Kamihara2008, Rotter2009}. 
While \textit{ab initio} density functional theory (DFT)
calculations can provide a qualitative understanding of their bandstructure and Fermi surface\cite{Lebegue2007,Singh2008,Mazin2008}, 
it became soon evident that correlation effects originating from the strong local Coulomb
repulsion on the Fe atoms are responsible for many experimental findings  such as large effective masses, Fermi surface
renormalization, finite lifetimes or transfer of spectral weight to high binding energies
\cite{Cao2008,Si2008,qazilbash2009,rullier2009,Haule2009,Medici2009,Aichhorn2010,Hansmann2010,Yin2011,Ferber2012,werner2012,Ferber_PRL_2012,hubbard2016,bascones2016,Watson2017}. 
The combined DFT with Dynamical Mean Field Theory (DFT+DMFT) method, which approximates the electronic self-energy
to be local in space and thus includes frequency- and orbital-dependent local effects of electronic correlations, has been very
successful in capturing many of these observations.  Some examples are orbital-dependent correlations, incoherence properties and
Fermi surface renormalization.\cite{Medici2009,Haule2009,Aichhorn2010,Hansmann2010,Yin2011,Ferber2012,werner2012,Ferber_PRL_2012,hubbard2016,Miao2016,Nourafkan2016,Watson2017}
However, the single-site DMFT cannot account for possible 
momentum-dependent correlation effects such as relative band shifts in opposite directions of, respectively, 
hole bands centered at $\Gamma$ and electron bands centered at the Brillouin 
zone edge $M$  (the so-called
``blue/red shift") in a large class of iron-based superconductors\cite{Ortenzi2009,Borisenko2010,Lee2012,Brouet2016,Kushnirenko2017}, or 
the recently reported\cite{Fink2018} possible momentum-dependent scattering rates in 
angular-resolved photoemission spectroscopy (ARPES) measurements of LiFeAs.
Some of these effects have been suggested to play an important role in the superconducting pairing mechanism~\cite{Borisenko2010,Gaplifeas2013,Miao2015,Altenfeld2018} as well.

Consideration of momentum dependence in the self-energy in real materials' calculations are scarce
but promising,~\cite{Biermann2003,Tomczak2012,Tomczak2014,Roekeghem2014,Biermann2014,LiKaushal2016,Semon2017} showing, for instance,
effects of bandwidth widening and momentum-dependent bandshifts in the systems studied~\cite{Tomczak2012,Roekeghem2014,LiKaushal2016}.
Here we explore this dependence by considering an approach where spin fluctuations play the
dominant role and it allows both, a description of local and non-local correlations on an equal footing.

The purpose of this work is twofold: (i) We first introduce the multi-orbital formulation of the Two-Particle Self-Consistent
(TPSC) approach originally conceived for the single-orbital Hubbard model~\cite{Vilk1997}, which provides
momentum- and frequency- dependent self-energies in the intermediate coupling regime. 
(ii) We apply the method to the iron-based superconductor LiFeAs.

We find that the momentum-dependence obtained within the TPSC approach introduces drastic 
changes to the LiFeAs Fermi surface and bandstructure with respect to DFT 
results.  First, the innermost hole pocket centered at $\Gamma$ is shifted below
the Fermi energy $E_F$ as de Haas-van Alphen (dHvA) and ARPES~\cite{Putzke2012,Brouet2016} 
measurements already suggested. Second, we find a large accumulation of 
incoherent spectral weight around the $\Gamma$ point as observed in 
ARPES~\cite{Borisenko2010,Lee2012,Miao2015,Brouet2016,Fink2018}. Third, the relative "blue/red shift" of the bands centered at 
$\Gamma$ and $M$ respectively\cite{Borisenko2010,Lee2012}, is properly described and, fourth, the momentum-averaged TPSC results
agree with the results obtained from previous local DFT+DMFT calculations~\cite{Ferber2012,Lee2012} pointing to an important relation 
between both approaches in this region of interactions.

{\it Models and methods.-} Starting from a DFT calculation of LiFeAs in the tetragonal crystal structure\cite{Morozov2010} 
within the Generalized Gradient Approximation (GGA)\cite{Perdew1996} using the full-potential linear augmented 
plane-wave basis from WIEN2K~\cite{Blaha2001}, we derive an effective low-energy
model comprizing the Fe $3d$ orbitals using maximally localized Wannier functions
as implemented in Wannier90\cite{Wannier90} (see Supplemental Material\cite{suppl}). 
We effectively then solve a 2-dimensional system by restricting our calculation to the $k_z=0$ plane,
since the low-energy electronic structure shows only weak dispersion along $k_z$. 
In this Wannier-projected 2D model we have an electron occupation of 6.
Interaction parameters for the lattice Hubbard model were obtained within the
constrained random-phase approximation (cRPA)\cite{fhigap} on the DFT bandstructure (see 
Supplemental Material\cite{suppl}).

The TPSC method considers the Luttinger-Ward functional $\Phi[G]$\cite{Martin1959,Baym1962},
which is a functional of the interacting Green's function $G$
and yields the self-energy $\Sigma$ and two-particle irreducible four-point vertex $\Gamma$ as functional
derivatives
\begin{align}
 \Sigma = \frac{\delta \Phi}{\delta G}, \ \ \ \ \Gamma &= \frac{\delta^2 \Phi}{\delta G^2}.
\end{align}
Within the TPSC approach one approximates the vertex $\Gamma$ to be static and momentum independent\cite{Vilk1997} (but fully orbital dependent).
One obtains a set of self-consistent and conserving equations that satisfy the Pauli principle 
and Mermin-Wagner theorem. 
The range of validity of TPSC is the regime of 
weak to intermediate couplings where the local and static approximation of the vertex is valid, i.e. away from any phase transition.
This method has been extended to 
multi-site,\cite{Arya2015,Aizawa2015,Zantout2018,Mertz2018,Aizawa2018,
Nishiguchi2018} nearest-neighbor interaction,\cite{Davoudi2006} and 
multi-orbital\cite{Miyahara2013} generalizations of the Hubbard model and has 
provided valuable insights on the pseudogap physics in the 
cuprates~\cite{Tremblay2012} and unconventional 
superconductivity.~\cite{Kyung2003,Ogura2015,Zantout2018}

In the multi-orbital generalization of the TPSC method 
similar to the original formulation\cite{Miyahara2013}
we first introduce the non-interacting susceptibility $\chi^0$
given by 
\begin{equation}
	\begin{array}{rl}
	\chi^0_{\lambda\mu\nu\xi}(\vec q, iq_m) = \left[G_{\nu\lambda}^0\star G^0_{\mu\xi}\right](\vec q,iq_m)
	\end{array}
	\label{eq:nonintsuscep}
\end{equation}
where $G^0$ denotes the non-interacting Green's function in orbital-space, $\star$ denotes a convolution over frequency
and momentum and $q_m = 2m\pi T$ the $m$-th bosonic Matsubara frequency.
The interacting susceptibility $\chi$ is decomposed into the spin and charge channel ( $\chi^\text{sp}$ and $\chi^\text{ch}$ respectively) and reads
\begin{equation}
	\begin{array}{rl}
		\chi^\text{sp}(\vec q, iq_m) &= [\mathbb{I}-\chi^{0}(\vec q,iq_m)U^\text{sp}]^{-1}2\chi^0(\vec q,iq_m)\\
		\chi^\text{ch}(\vec q, iq_m) &= [\mathbb{I}+\chi^{0}(\vec q,iq_m)U^\text{ch}]^{-1}2\chi^0(\vec q,iq_m),
	\end{array}
	\label{eq:chi_sp/ch}
\end{equation}
where the inversion of a 4-index tensor is given as the matrix inverse
after combining the first and last two indices of $\lambda\mu\nu\xi$ 
into a superindex $(\lambda\mu)(\nu\xi)$.

We only consider the $U^\text{ch/sp}_{\alpha\alpha\beta\beta}$ 
and $U^\text{ch/sp}_{\alpha\beta\alpha\beta}=U^\text{ch/sp}_{\alpha\beta\beta\alpha}$
matrix elements of the renormalized irreducible vertices in the spin $U^\text{sp}$ and 
the charge channel $U^\text{ch}$ to be nonzero, corresponding to the atomic symmetry of $3d$ orbitals.
Those elements are determined by enforcing the following local spin and charge sum rules
\begin{equation}
	\begin{array}{rl}
		&\frac{T}{N_{\vec q}}\sum\limits_{\vec q,m}\chi^\text{sp}_{\mu\nu\mu\nu}(\vec q, iq_m) = \langle n_{\mu\uparrow}\rangle+\langle n_{\nu\uparrow}\rangle - 2\langle n_{\mu\uparrow}n_{\nu\downarrow}\rangle,\\
		&\frac{T}{N_{\vec q}}\sum\limits_{\vec q,m}\chi^\text{sp}_{\mu\mu\nu\nu}(\vec q, iq_m) \stackrel{\mu\neq\nu}{=} 2\langle n_{\mu\uparrow}n_{\nu\uparrow}\rangle - 2\langle n_{\mu\uparrow}n_{\nu\downarrow}\rangle,\\
		&\frac{T}{N_{\vec q}}\sum\limits_{\vec q,m}\chi^\text{ch}_{\mu\mu\nu\nu}(\vec q, iq_m) = 2\langle (n_{\mu\uparrow}+n_{\mu\downarrow})n_{\nu\uparrow}\rangle - n_\mu n_\nu,\\
		&\frac{T}{N_{\vec q}}\sum\limits_{\vec q,m}\chi^\text{ch}_{\mu\nu\mu\nu}(\vec q,iq_m) \stackrel{\mu\neq\nu}{=} \frac{n_{\mu}+n_{\nu}}{2}-\langle (4n_{\mu\uparrow}-2n_{\mu\downarrow})n_{\nu\uparrow}\rangle.
	\end{array}
\label{eq:sumrules}
\end{equation}
In order to solve this underdetermined set of equations we employ an ansatz for the spin 
vertex $U^\text{sp}$ that is motivated by the Kanamori-Brueckner 
screening~\cite{Vilk1997,Miyahara2013} and introduce an additional particle-hole symmetrization to keep all equations within TPSC particle-hole symmetric: 
\begin{equation}
	\begin{array}{rl}
		U^\text{sp}_{\mu\mu\mu\mu} &= \frac{1}{2}\left(\frac{\langle n_{\mu\uparrow}n_{\mu\downarrow}\rangle}{\langle n_{\mu\uparrow}\rangle \langle n_{\mu\downarrow}\rangle} + \text{particle$\leftrightarrow$hole}\right)U_{\mu\mu}\\
		U^\text{sp}_{\mu\nu\mu\nu} &= \frac{1}{2}\left[\frac{\langle n_{\mu\uparrow}n_{\nu\downarrow}\rangle}{\langle n_{\mu\uparrow}\rangle \langle n_{\nu\downarrow}\rangle}U_{\mu\nu} + \frac{\langle n_{\mu\uparrow}n_{\nu\uparrow}\rangle}{\langle n_{\mu\uparrow}\rangle \langle n_{\nu\uparrow}\rangle}(U_{\mu\nu}-J_{\mu\nu})\right.\\
		&\left.~~~~~+ \text{particle$\leftrightarrow$hole}\right]=U^{\text{sp}}_{\mu\mu\nu\nu}=U^{\text{sp}}_{\mu\nu\nu\mu}.
		\end{array}
\label{eq:Uspansatz}
\end{equation}
The local spin vertex $U^\text{sp}$ can be obtained by iterating the 
equations above.
For the charge channel we optimize $U^{\text{ch}}$ in order to fulfill
the corresponding charge sum rules, restricting to positive values of $U^{\text{ch}}$
because in certain cases negative values can lead to non-causal self-energies.
Due to the constraint we search for values of $U^{ch}$ that minimize (see Supplemental Material~\cite{suppl}) the difference between left-hand side and
right-hand side of the charge sume rules (Eq.~\eqref{eq:sumrules}).

After the determination of the spin and charge vertices the self-energy $\Sigma$ and interacting Green's function $G$ are then
given as
\begin{align}
	\Sigma_{\mu\nu} =& \frac{1}{4}\sum\limits_{\alpha,\beta}\underbrace{\left[U^\text{sp}\chi^\text{sp}U^\text{sp,0}+U^\text{ch}\chi^\text{ch}U^\text{ch,0}\right]_{\nu\alpha\mu\beta}}_{:=V_{\nu\alpha\mu\beta}}\star G^0_{\beta\alpha}\nonumber\\
	G(\vec k, i\omega_n) =& \left[(i\omega_n + \mu)\mathbb{I} - H_\text{0}(\vec k) -\Sigma(\vec k, i\omega_n)\right]^{-1},\label{eq:GSigma}
\end{align}
where the non-interacting vertices are zero except for the matrix elements: 
$U^\text{sp/ch,0}_{\mu\mu\mu\mu}=U_{\mu\mu}$, $U^\text{ch,0}_{\mu\mu\nu\nu}=2U_{\mu\nu}-J_{\mu\nu}$ and $U^\text{sp/ch,0}_{\mu\nu\mu\nu}=U^\text{sp/ch,0}_{\mu\nu\nu\mu}=U^\text{sp,0}_{ \mu\mu\nu\nu}=J_{\mu\nu}$ with $\mu\neq\nu$. 
No Hartree term is included in $\Sigma$ since it is already contained 
in the DFT-derived Hamiltonian $H_0$.

Our multi-orbital extension of TPSC differs from previous formulations~\cite{Miyahara2013} on the following aspects:
it restricts the self-consistent equations  in the charge channel in Eq.~\eqref{eq:chi_sp/ch} to ensure positivity of the spectral weight and chooses a symmetrized ansatz for $U^{sp}$ (Eq.~\eqref{eq:Uspansatz}).
Furthermore, the set of local spin and charge sum rules (Eq.~\eqref{eq:sumrules}) 
and bare vertex tensors $U^{ch,0}, U^{sp,0}$ (Eq.~\eqref{eq:GSigma}) and their dependence on $U, J$ are derived from the Bethe-Salpeter equation for the self-energy $\Sigma$ within TPSC~\cite{Vilk1997}, which is different from the RPA derived expression of $U^{ch,0}, U^{sp,0}$ only in the $ijij$-element, $i\neq j$.\cite{Miyahara2013}

Our calculations were performed at T=0.015 eV$\approx$174 K since this is the 
lowest accessible temperature before spin fluctuations get too strong and the 
TPSC approximation is not justified anymore. Nevertheless, we checked that the 
results presented below do not change in their trends up to room 
temperature (see Supplemental Material~\cite{suppl}).
\begin{figure}[h]
	\includegraphics[width=0.9\linewidth]{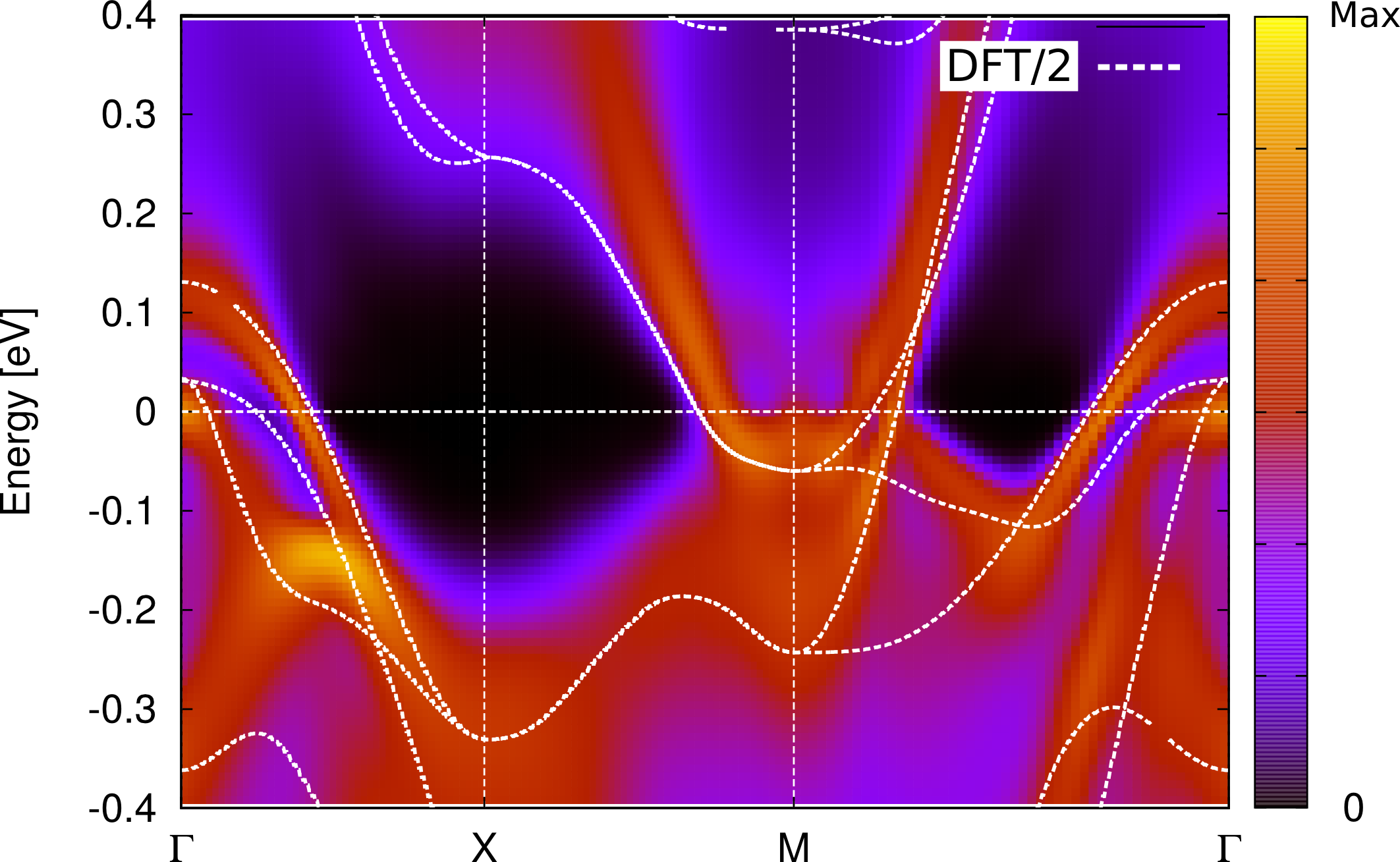}
	\caption{Interacting spectral function $A(\vec k, \omega)$ within the TPSC approach 
	for LiFeAs in the two-iron Brillouin zone. 
	For comparison we show the DFT(GGA) bandstructure renormalized by the average mass 
        enhancement $\approx 2$ (dotted lines).
	We observe an overall shrinking of the electron and hole pockets at $\Gamma$ and $M$ originating
	from the non-local components of the self-energy.
	The center hole pockets at $\Gamma$ become incoherent and diffuse due to finite lifetime effects 
	in the Fe $3d_{xz}/3d_{yz}$ orbitals.}
	\label{fig:bandstructure_LiFeAs}
\end{figure}

\begin{figure*}
	\includegraphics[width=\textwidth]{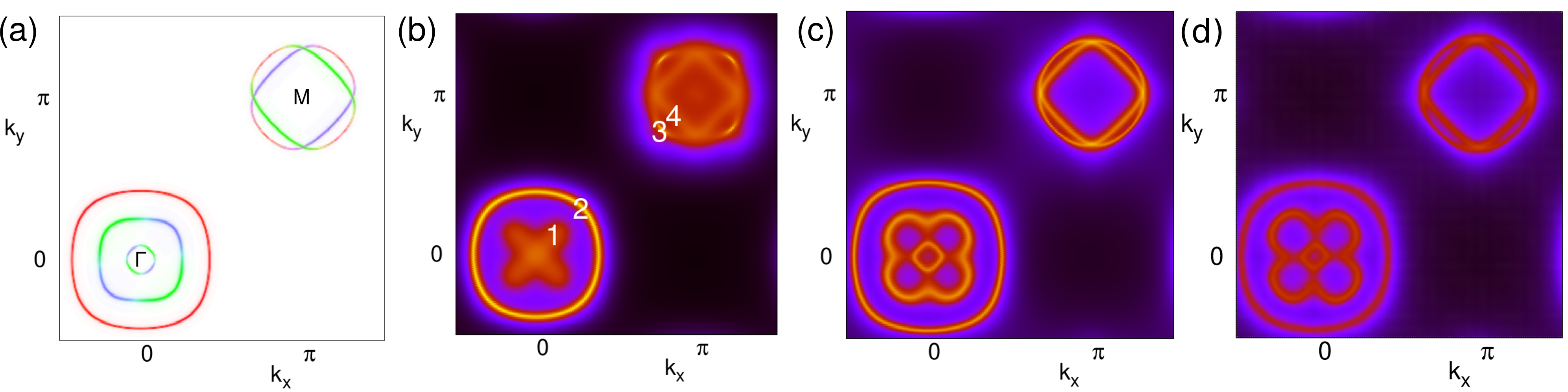}
	\caption{(a) Orbital-resolved Fermi surface obtained from DFT(GGA) where the dominant  orbital 
	characters are $d_{xy}$ (red), $d_{yz}$ (blue) and $d_{xz}$ (green). 
	 Three hole pockets are centered around $\Gamma$ and two electron 
	 pockets around the $M$ point. (b) Fermi surface from DFT+TPSC. 
	 We observe strong incoherence effects on the inner hole and electron pockets.
         The two inner hole pockets become very incoherent 
	 and form a flower-like shaped region of spectral weight. 
	 (c) Fermi surface from DFT+"local TPSC" where the momentum dependent TPSC self-energy 
	 $\Sigma(\vec k,\omega)$ has been approximated by its local 
	 component $\frac{1}{N_{\vec k}}\sum_{\vec k}\Sigma(\vec k,\omega)$. 
	 In this approximation the Fermi surface recovers well 
	 published DFT+DMFT result~\cite{Ferber2012,Lee2012}. (d) DFT+DMFT Fermi surface for the same model (see Supplemental Material~\cite{suppl}) as used in this work. We see a strong similiarity to the DFT+"local TPSC" result in (c).}
	\label{fig:Fermi_Surface_LiFeAs_GGA}
\end{figure*}

{\it Results and discussion.-}
In Fig.~\ref{fig:bandstructure_LiFeAs} we show the TPSC spectral function $A(\vec k, 
\omega)$ for LiFeAs along $\Gamma-X-M-\Gamma$ in the two-iron Brillouin 
zone. To emphasize the changes in the electronic
structure beyond an overall bandwidth renormalization of about a factor of $2$,
we also plot the renormalized DFT bandstructure on top. 
We observe that the electronic correlations introduce a downshift
of the hole states around the Fermi level at the $\Gamma$-point, while
the electron states at $M$ are slightly shifted up in energy. The inner electron pocket being shifted
by -0.1eV on average while the outer electron pocket is shifted by only -0.01eV.
This leads to an overall shrinking of hole and electron pockets,
corresponding to the ``blue/red shift'' seen in ARPES measurements\cite{Borisenko2010,Brouet2016}
compared to the DFT bandstructure.
Apart from orbital dependence the shrinkings are momentum-dependent.
For example, along $\Gamma-X$ the middle hole pocket
shrinks to approximately 20\% of its size compared to the renormalized DFT bandstructure while all other
parts of the Fermi surface shrink to 80-90\% of their original size.
The inner hole pocket at $\Gamma$, composed of Fe  $3d_{xz}/3d_{yz}$ orbital character (see Fig.~\ref{fig:Fermi_Surface_LiFeAs_GGA} (a)),
becomes very diffuse at the Fermi level due to incoherent scattering processes,
leading to a significant reduction of the lifetime of quasi-particle excitations.
This manifests in a broad Fermi surface feature very similar to the one observed in
ARPES\cite{Borisenko2010,Lee2012,Miao2015,Brouet2016,Fink2018}.
The maximum of the spectral function of the two inner hole pockets at $\Gamma$
is shifted basically on top of the Fermi level but retains significant
spectral weight at higher and lower binding energies - the shift being on average 0.18eV for both while for the outer hole pocket it is 0.1eV.
We expect that the inclusion of spin-orbit coupling, which is beyond our current approach,
will split this feature, effectively pushing one hole band below and the other 
above the Fermi level, giving rise to only one central hole Fermi surface pocket,
which would be in very good agreement with previous
ARPES data~\cite{Brouet2016, Kushnirenko2018} as well as de Haas-van 
Alphen (dHvA) experiments.\cite{Putzke2012}

We can trace back these Fermi surface modifications
to the value of the self-energy at the specific $\vec k$-points in the Brillouin zone:
The largest contribution to the diagonal elements of 
the self-energy in Eq.~\eqref{eq:GSigma} stems from $V_{abab}$,
which is peaked at  $\vec k=\{ (\pm\pi,0) , (0,\pm\pi) \}$. Following 
the argumentation of Ref.~\onlinecite{Ortenzi2009}, this leads to 
a negative (positive) real part of the self-energy in the vicinity of the hole (electron)
pockets and thus to the observed ``blue/red shift'' and therefore it is
a consequence of non-local 
spin fluctuations.

In Fig.~\ref{fig:Fermi_Surface_LiFeAs_GGA} we show the orbitally-resolved Fermi surface obtained from DFT (within GGA) 
(Fig.~\ref{fig:Fermi_Surface_LiFeAs_GGA} (a)), DFT+TPSC 
(Fig.~\ref{fig:Fermi_Surface_LiFeAs_GGA} (b)) and DFT+``local TPSC'' where the 
momentum dependent TPSC self-energy $\Sigma(\vec k,\omega)$ has 
been approximated by its local component 
$\frac{1}{N_{\vec k}}\sum_{\vec k}\Sigma(\vec k,\omega)$(Fig.~\ref{fig:Fermi_Surface_LiFeAs_GGA} (c)). 
The DFT Fermi surface reveals 
three well-defined distinct hole pockets centered at $\Gamma$  with circular to square shape and two 
electron pockets centered at $M$. As can already be deduced from the spectral 
function $A(\vec k,\omega)$ in Fig.~\ref{fig:bandstructure_LiFeAs}, the Fermi 
surface experiences appreciable changes due to the TPSC self-energy 
contributions. All pockets are reduced in size,
with the remaining spectral weight of the two center hole pockets
of Fe $3d_{xz}/3d_{yz}$ character at 
$\Gamma$ becoming incoherent and forming a 
flower-like shape, while the outer hole pocket of $3d_{xy}$ character stays coherent as confirmed in ARPES measurements~\cite{Borisenko2010,Brouet2016}.
The electron pockets at $M$ shrink slightly and broaden, since they are mostly composed
of the most incoherent $3d_{xz}/3d_{yz}$ as well.

The observed shrinking of the hole and electron pockets deviates significantly
from published DFT+DMFT results,
most likely due to the inclusion of non-local correlations in the TPSC approach
which go beyond the DMFT approximation where the self-energy is purely local.
In order to confirm this assumption we separate the local 
from the non-local correlation effects by employing 
a DMFT-like approximation on the TPSC self-energy. We approximate the full
momentum-dependent TPSC self-energy $\Sigma(\vec k, \omega)$ 
by its local component and compare the resulting Fermi surface to
the full result in Fig.~\ref{fig:Fermi_Surface_LiFeAs_GGA} (c). 
The so obtained Fermi surface indeed recovers the result obtained within published DFT+DMFT~\cite{Ferber2012,Lee2012}, and, when considering the DFT+DMFT results for the same model as used in this work (see Supplemental Material~\cite{suppl}), the agreement is almost perfect 
(compare Fig.~\ref{fig:Fermi_Surface_LiFeAs_GGA} (c) and (d)). DFT+DMFT calculations with a different double counting scheme~\cite{Nourafkan2016} see a more pronounced -although coherent- flower-like shape of spectral weight around $\Gamma$ but don't account for the ``blue/red shift''.
This shows that when taking into account non-local fluctuations,
the local Coulomb interaction gives rise to a significant momentum-dependent self-energy
and can account for the experimentally observed ``blue/red shift''.
Interestingly, within the local approximation (local TPSC) the center hole pockets at $\Gamma$ become again coherent, 
which is also in correspondence with the DMFT result. 
This shows that the quasi-particle scattering rate itself is strongly
momentum and orbital dependent, which has in fact been observed in
recent ARPES experiments\cite{Brouet2016,Fink2018},
where the inner $3d_{xz}/3d_{yz}$ derived hole Fermi surface have been found to be incoherent
while the outer $3d_{xy}$ hole pocket shows Fermi liquid behavior.

\begin{figure}[h]
	\includegraphics[width=0.8\linewidth]{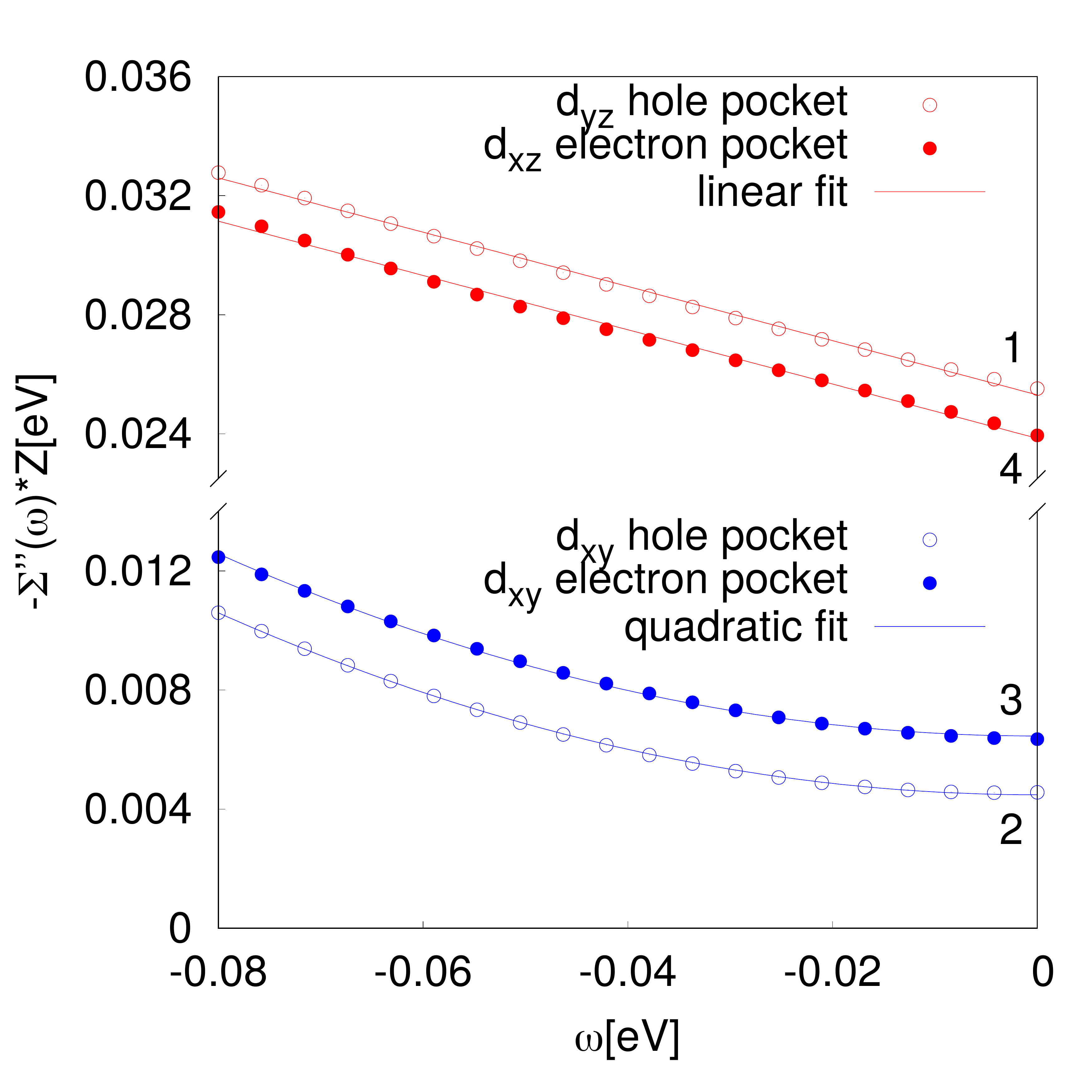}
	\caption{Quasi-particle lifetimes $-Z_{\vec k}\Sigma''(\vec
	k,\omega)$ along $\Gamma-M$ as a function of the binding energy $\omega$ (numbers on the right correspond to positions in Fig.\ref{fig:Fermi_Surface_LiFeAs_GGA}(c)).	We find that the quasi-particles with $d_{xz/yz}$ character	display a  linear dependence in $\omega$ while the electron pockets have a quadratic increase with energy. (See Supplemental Material~\cite{suppl})}
	\label{fig: ImSigma}
\end{figure}

Since Fermi-liquid theory predicts a quadratic energy dependence of quasi-particles' 
lifetimes near the Fermi energy, deviations from this energy dependence  
are also a signature for non-Fermi liquid behavior. It is 
therefore compelling to analyze the energy dependencies of the scattering rate within the
TPSC approach. 
For this, we present the quasi-particle lifetime $-Z_{\vec k}\Sigma''(\vec k,w)$ with
\begin{equation}
	Z_{\vec k}=\frac{1}{1-\frac{\partial \Sigma''(\vec k,i\omega_n)}{\partial w_n}|_{i \omega_n\rightarrow0^+}}
\end{equation}
in Fig.~\ref{fig: ImSigma} at four different $\vec k$-points in momentum space following 
the  $\Gamma-M$ path. Along this path the dominating contributions  are (1) $d_{yz}$
hole pocket, (2) $d_{xy}$ hole pocket, (3) $d_{xy}$ electron pocket and (4) 
$d_{xz}$ electron pocket (see Fig.~\ref{fig:Fermi_Surface_LiFeAs_GGA}(a) and 
(b)).  The energy dependence of the quasi-particle lifetimes  for the 
$d_{xz/yz}$ electron and hole pockets (red symbols in Fig.~\ref{fig: ImSigma}) 
are in good agreement with the results of Ref.~\onlinecite{Brouet2016} with 
values between 0.025eV and 0.035eV. The energy 
dependence shows a very shallow linear behavior (fitted slopes are of the order of $10^{-3}$, see Supplemental Material~\cite{suppl}) similar to the measurements from Ref.~\onlinecite{Brouet2016}. The  quasi-particle lifetimes of the 
$d_{xy}$ hole and electron pockets (blue symbols in Fig.~\ref{fig: ImSigma}), in 
contrast, show at the considered $k$-points a quadratic increase in energy as in 
the ARPES measurements of Ref.~\onlinecite{Brouet2016}, suggesting a 
Fermi-liquid-like behavior. Although our data was obtained at
$T\approx 174\text{K}$ in contrast to the $T=25\text{K}$ in
Ref.~\onlinecite{Brouet2016}, we are confident that our results are still valid at low temperatures, 
since for example in
Ba(Fe$_{0.92}$Co$_{0.08}$)$_2$As$_2$ it has been found that the quasi-particle lifetimes for the hole 
$d_{xz/yz}$ orbitals showed weak temperature dependence. We also checked how these 
results depend on the $\vec k$-path and found that small translations along the tip of electron pocket (3) reveal a linear dependence of the quasi-particle lifetime as can already be expected since the quasi-particle 
weight gets incoherent away from the point (3) (see 
Fig.~\ref{fig:Fermi_Surface_LiFeAs_GGA}(b)).

{\it Summary.-} In conclusion, we presented a multi-orbital TPSC scheme that 
respects local spin and charge sum rules. This method includes effects
of local and non-local correlations on an equal footing 
within the validity of the local approximation of the 
irreducible 4-point vertex and thus yields momentum- and 
frequency-dependent self-energies. 
We applied this method to the multi-orbital iron-based superconductor LiFeAs
and found that the non-local components of the self-energy are decisive to 
explain its experimentally observed spectral function $A(k,\omega)$ and Fermi surface.
Taking into account non-local correlations we observe 
a ``blue/red shift" of the electronic structure, where the hole bands 
at the Brillouin zone center are lowered in energy, while the electron bands 
in the corner of the Brillouin zone are slightly shifted upwards, resulting
in an overall reduction of the size of the Fermi surface pockets.
Overall we find very good agreement with ARPES and dHvA experiments,
where the ``blue/red shift" was first observed.
We could show that our TPSC approach within a local approximation to the self-energy
recovers the DFT+DMFT result which does not exhibit 
the ``blue/red shift", both benchmarking 
 the TPSC result and showing the importance of going beyond the local
picture of DMFT in order to understand the electronic structure of
iron-based superconductors.
Furthermore, we also found a strong momentum and non-quadratic energy dependence
of the electronic scattering rate, in good agreement with recent ARPES measurements.

\begin{acknowledgments}
This work was supported by the German Research Foundation (Deutsche 
Forschungsgemeinschaft) under grant SFB/TR 49. We would like to thank S. 
Bhattacharyya, S. Borisenko, V. Borisov, J. Fink, P.J. Hirschfeld, G. Kotliar and Y. Li for discussions.
R. V. acknowledges the KITP for hospitality
where aspects of this work were discussed.  The research
at KITP was supported in part by NSF Grant No. NSF
PHY-1748958.
\end{acknowledgments}

\newpage
\bibliographystyle{apsrev4-1}
\bibliography{Literatur_paper.bib}
\clearpage
\newpage


\section*{Supplementary material}

# Supplemental Material: Effect of non-local correlations on the electronic structure of LiFeAs

Karim Zantout,[1] Steffen Backes,[2] and Roser Valentí[1]

[1]*Institut für Theoretische Physik, Goethe-Universität Frankfurt, 60438 Frankfurt am Main, Germany*
[2]*CPHT, CNRS, Institut Polytechnique de Paris, F-91128 Palaiseau, France*

## I. LOW-ENERGY MODEL

Our DFT-derived low-energy Hamiltonian is given by

$$H = \underbrace{\sum_{\alpha,\beta,i,j,\sigma} (t_{ij,\alpha\beta} - \mu \delta_{i,j}\delta_{\alpha,\beta}) c^\dagger_{i\alpha\sigma} c_{j\beta\sigma}}_{:=H_0}$$

$$+ \sum_{\alpha,\beta,i,\sigma} \frac{U_{\alpha\beta}}{2} n_{i\alpha\sigma} n_{i\beta-\sigma} + \sum_{\substack{\alpha,\beta,i,\sigma \\ \alpha\neq\beta}} \frac{U_{\alpha\beta} - J_{\alpha\beta}}{2} n_{i\alpha\sigma} n_{i\beta\sigma}$$

$$- \sum_{\substack{\alpha,\beta,i,\sigma \\ \alpha\neq\beta}} \frac{J_{\alpha\beta}}{2} \left( c^\dagger_{i\alpha\sigma} c_{i\alpha-\sigma} c^\dagger_{i\beta-\sigma} c_{i\beta\sigma} + c^\dagger_{i\alpha\sigma} c_{i\beta-\sigma} c^\dagger_{i\alpha-\sigma} c_{i\beta\sigma} \right),$$

where $t_{ij,\alpha\beta}$ denote the hopping parameters from orbital $\alpha$ in unit cell $i$ to orbital $\beta$ in unit cell $j$, $\mu$ is the chemical potential, $c^\dagger_{i\alpha\sigma}$ ($c_{i\alpha\sigma}$) creates (annihilates) an electron in unit cell $i$ at orbital $\alpha$ with spin $\sigma$. The Hamiltonian is constructed in the basis of the Fe 3$d$ orbitals using maximally localized Wannier functions as implemented in Wannier90[44].

## II. INTERACTION PARAMETERS

The interaction parameters obtained from cRPA are given as (in the $d_{z^2}$, $d_{x^2-y^2}$, $d_{xy}$, $d_{xz}$, $d_{yz}$ basis of the Fe 3d orbitals)

$$U = \begin{pmatrix} 3.40 & 1.94 & 2.03 & 2.39 & 2.39 \\ 1.94 & 2.54 & 2.13 & 1.89 & 1.89 \\ 2.03 & 2.13 & 2.66 & 1.98 & 1.98 \\ 2.39 & 1.89 & 1.98 & 2.75 & 2.02 \\ 2.39 & 1.89 & 1.98 & 2.02 & 2.75 \end{pmatrix} \text{eV}$$

and

$$J = \begin{pmatrix} 0 & 0.49 & 0.51 & 0.34 & 0.34 \\ 0.49 & 0 & 0.23 & 0.39 & 0.39 \\ 0.51 & 0.23 & 0 & 0.41 & 0.41 \\ 0.34 & 0.39 & 0.41 & 0 & 0.41 \\ 0.34 & 0.39 & 0.41 & 0.41 & 0 \end{pmatrix} \text{eV}. \quad (1)$$

We employed a $d-d$ model, i.e. the same Wannier basis was used for the Hamiltonian and the screening of only the Fe 3d orbitals has been removed. A k-mesh of $8 \times 8 \times 5$ was used for the integration over the Brillouin zone.

Compared to previously published interaction values of LiFeAs obtained from cRPA[61] we observe smaller values of $U_{d_{xy},d_{xy}} = 2.66$eV instead of 3.51eV or $U_{d_{xz},d_{xz}} = U_{d_{yz},d_{yz}} = 2.75$eV instead of 3.16eV. We attribute the differences to the following points:

1) In Ref. 61 the DFT calculation has been performed using the full-potential LMTO basis, while we have used the full-potential LAPW basis from Wien2K[43].

2) We found that the cRPA calculation is sensitive to the resolution of the discretization in momentum space and a fine grid is needed for proper convergence. In our calculation we employed an 8x8x5 k-mesh, while the reference used a smaller mesh of 4x4x4. We checked the dependence on the k-mesh by performing the calculation with the smaller meshes and found a significant difference in the spread of the Wannier functions:

|  | Spread of $d_{z^2}$ | $d_{x^2-y^2}$ | $d_{xy}$ | $d_{xz/yz}$ |
|---|---|---|---|---|
| 4x4x4 k-points | 2.85Å$^2$ | 3.53Å$^2$ | 2.85Å$^2$ | 3.09Å$^2$ |
| 8x8x5 k-points | 2.91Å$^2$ | 8.74Å$^2$ | 3.84Å$^2$ | 3.91Å$^2$ |

The spread increases significantly for a denser k-mesh, explaining why we find lower interaction values with an 8x8x5 k-mesh compared to the 4x4x4 mesh used in Ref. 61. We also performed the cRPA calculation for 4x4x4 k-meshes and obtained larger interaction values that are closer to the one shown in the reference: $U_{d_{xy},d_{xy}} = 3.03$eV or $U_{d_{xz},d_{xz}} = U_{d_{yz},d_{yz}} = 2.8$eV. We think our results are sufficiently converged since the values changed only about 5-10% when increasing the mesh from 7x7x5 to 8x8x5 k-points. A finer calculation with a 9x9x6 k-mesh could not be done due to memory constraints.

## III. COMPUTATIONAL DETAILS

For the momentum-space integration we employed an adaptive cubature method based on a three-point formula for triangles with an integration tolerance of $10^{-6}$ to evaluate the non-interacting susceptibility. We obtained all quantities on a $100 \times 100$ $\vec{k}$-grid in order to account for any singular features. The fast Fourier transformation and the circular convolution theorem was used for an efficient implementation of Eq. 6 in the main document. The summation over fermionic Matsubara frequencies was performed for $N_{\text{Mats}} = 250 \cdot (0.025/T)$eV points and for bosonic Matsubara frequencies we took $2N_{\text{Mats}}/3$ points. In both cases we included high-frequency corrections up to the order of $\frac{1}{\omega^2}$ by extrapolation of the high-frequency tail. The calculations were performed at $T = 0.015 eV \approx 174$K. Analytical continuation was performed with the maximum entropy Code from[62] in the case of the spectral function $A(\vec{k}, \omega)$ and with Padé approximation for the imaginary part of the self-energy $\Sigma''(\vec{k}, \omega)$ and calculated $Z_{\vec{k}}$ via linear extrapolation $i\omega \to 0$ in the case of quasi-particle lifetimes.

During the optimization procedure of $U^{ch}$ we encounter deviations of at most 6% between left-hand and right-hand side of the charge sum rules (Eq. 4 in the main document) in our calculations of LiFeAs. Since the charge channel has a much smaller contribution to the self-energy $\Sigma$ compared to the spin channel we can estimate the effect of this inaccuracy on our results to be negligible.

The fitting curves $\tau$ for the quasi-particle lifetimes $-Z_{\vec{k}}\Sigma''(\vec{k},\omega)$ from Fig. 3 are given by

$$d_{yz} \text{ hole pocket: } \tau(\omega) = 25\text{meV} - 0.09\omega$$
$$d_{xz} \text{ electron pocket: } \tau(\omega) = 24\text{meV} - 0.091\omega$$
$$d_{xy} \text{ hole pocket: } \tau(\omega) = 4.5\text{meV} + \frac{0.95}{\text{eV}}\omega^2$$
$$d_{xz} \text{ electron pocket: } \tau(\omega) = 6.5\text{meV} + \frac{0.96}{\text{eV}}\omega^2.$$

## IV. TEMPERATURE DEPENDENCE

In the range of temperatures studied which is $T \gtrsim 175$K we see that the trends in the results stay the same, for instance the Fermi surface keeps its topology (Fig. 4). We also see that the quasi-particle lifetime $-Z_{\vec{k}}\Sigma''(\vec{k},\omega=0)$

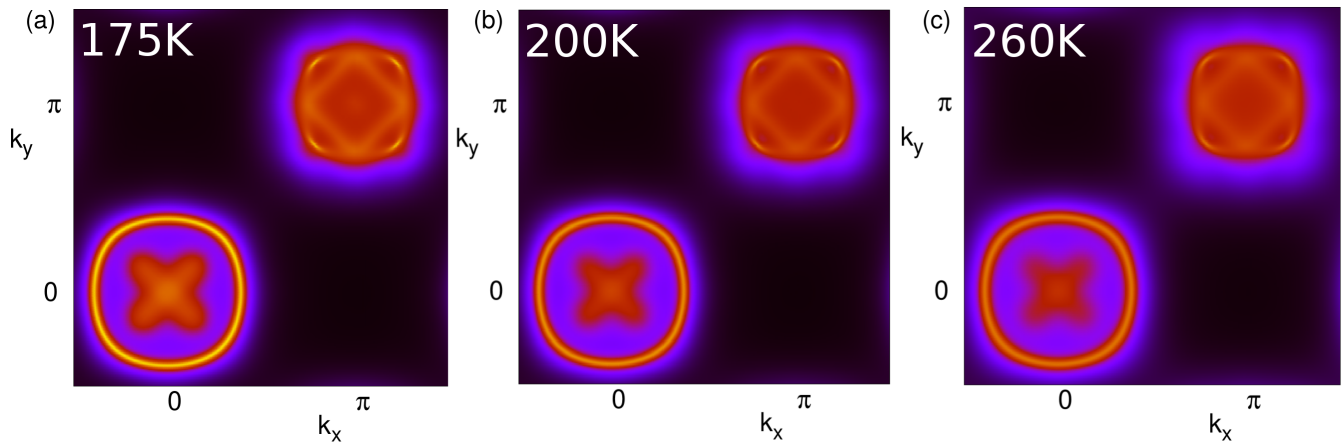

FIG. 1. DFT+TPSC Fermi surfaces at (a) $T = 175$ K, (b) $T = 200$ K and (c) $T = 260$ K. We see that the Fermi surface changes only marginally due to thermal decoherence.

at the $d_{xy}$ hole pocket seems to show a linear increase (see Fig. 5) in accordance with the coherence-incoherence crossover region measured in Ref. 21.

## V. LOCAL TPSC AND DMFT

We performed a local approximation to the TPSC self-energy and found that it recovers the published DMFT result[15,25]. This means the Fermi surface is almost identical to the DFT Fermi surface except for subtle changes in the middle hole pocket around $\Gamma$. This shows that the TPSC approximation is able to access the local correlations in LiFeAs in the same extent as DMFT employing a numerically exact impurity solver. Since our model involves additional approximations discussed in the main manuscript (two-dimensional lattice using a one iron unit cell) we performed an additional standard DMFT calculation with the same noninteracting Hamiltonian and identical parameters used in the TPSC calculation. For the solution of the impurity model we used the continuous-time Quantum Monte-Carlo Solver in the segment picture as implemented in the ALPSCore package[44,63].

In Fig.6 we show a comparison between the low-energy momentum-resolved spectral function obtained from the local TPSC and the DMFT calculation, compared to the renormalized DFT result. Both cases basically reproduce the DFT electronic structure at the Fermi level except for an overal bandwidth renormalization, which is slightly stronger in DMFT (factor 2.1) than in local-TPSC (factor 2), and stronger degree of broadening at higher energies. There is an additional small modification of the middle hole pocket along $\Gamma - M$ in DMFT, which is not captured in local TPSC. The fact that both approaches agree very well and very closely reproduce the DFT electronic structure, most importantly not exhibiting an overall shrinking of Fermisurface pockets with a blue/red shift, shows that TPSC is able to capture the local electronic correlations and that non-local extensions are needed to explain the electronic structure of LiFeAs.

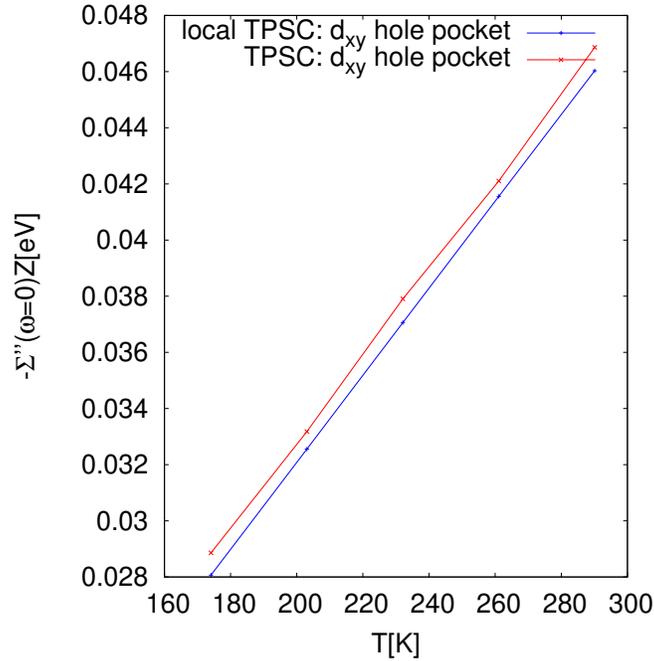

FIG. 2. Quasi-particle lifetime at the $d_{xy}$ hole pocket within DFT+TPSC and DFT+"local" TPSC. We see a linear increase in both due to thermal decoherence and a weak momentum-dependence.

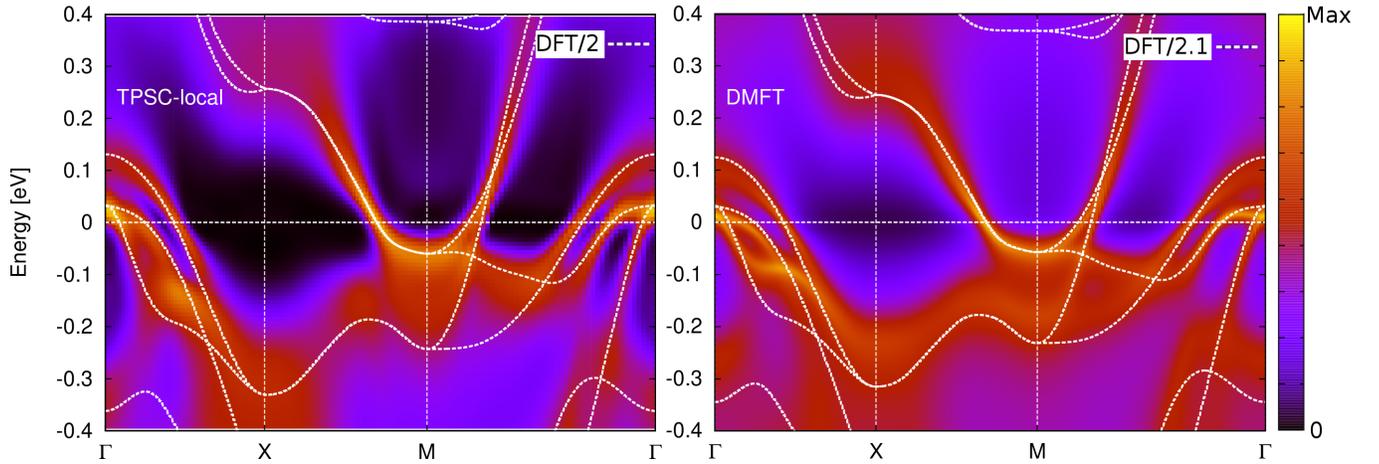

FIG. 3. Left: The momentum-resolved spectral function of LiFeAs within the local TPSC approximation (see main text). Right: DMFT The momentum-resolved spectral function using the same noninteracting Hamiltonian and parameters as in the TPSC calculation.

To quantitatively compare the two approaches, we plot a comparison of the local TPSC self-energy and the DMFT impurity self-energy in Fig.7. We observe a good agreement between the two approaches, especially in the low-energy part of the imaginary part of the self-energy. This part is the relevant one for the low-energy electronic states like the Fermi surface. A noticable difference is the stronger renormalization of about a factor of 2.1 on average in DMFT compared to TPSC with a factor of 2, resulting in a steeper slope of the self-energy close to $\omega = 0$ eV. For higher energies the difference between the two approaches becomes much larger, an effect that has already been observed in the single-band TPSC[39] where it was found that TPSC in general underestimates the tail of the self-energy at higher energies, but obtains accurate results for lower energies. Since the local self-energy is the only correction to the electronic structure in the DMFT and local TPSC approximation, this shows that the local TPSC approach is able to capture the local electronic correlations to a high degree.

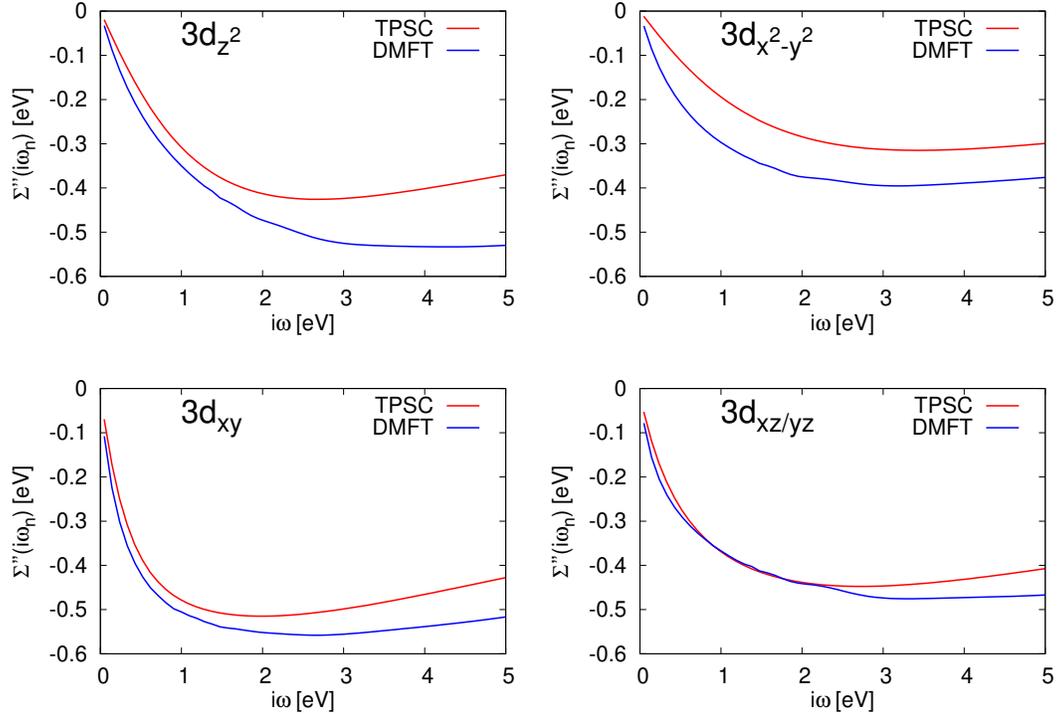

FIG. 4. The imaginary part of the local TPSC self-energy compared to the impurity self-energy obtained from the DMFT approximation. We observe a good agreement in the low-energy part (except for the $d_{x^2-y^2}$ orbital), which is the most relevant for the low-energy electronic structure like the Fermi surface.


\end{document}